\documentclass[12pt]{article}
\def\slash#1{\setbox0=\hbox{$#1$}#1\hskip-\wd0\hbox to\wd0{\hss\sl/\/\hss}}
\usepackage{epsf, cite, amsmath, amssymb}
\usepackage{graphicx}
\usepackage[all]{xy}
\makeatletter
\renewcommand\section{\@startsection {section}{1}{\z@}%
                                   {-3.5ex \@plus -1ex \@minus -.2ex}%nn
                                   {2.3ex \@plus.2ex}%
                                   {\normalfont\large\bfseries}}
\renewcommand\subsection{\@startsection{subsection}{2}{\z@}%
                                     {-3.25ex\@plus -1ex \@minus -.2ex}%
                                     {1.5ex \@plus .2ex}%
                                     {\normalfont\bfseries}}
\makeatother

\let\non\nonumber

\newcommand{\bea}{\begin{eqnarray}}
\newcommand{\eea}{\end{eqnarray}}
\newcommand{\be}{\begin{equation}}
\newcommand{\ee}{\end{equation}}

\def\sst#1{\mbox{\tiny#1}}

\newcommand{\p}{\partial}

\newcommand{\s}{\sigma}

\newcommand{\Co}{{\cal{O}}}

\newcommand{\C}[1]{$(\ref{#1})$}

\def\twoways{1}
\begin{document}
\begin{titlepage}

\begin{center}

\hfill EFI-05-15

\vskip 2 cm
{\Large \bf Boundary Ground Ring in Minimal String Theory}\\
\vskip 1.25 cm { Anirban 
Basu\footnote{email: basu@theory.uchicago.edu, \\Address 
after Sep 11, 2005: School of Natural Sciences, Institute for Advanced 
Study, Princeton, NJ 08540, USA}
and Emil J. Martinec\footnote{email: ejm@theory.uchicago.edu}
}\\
{\vskip 0.75cm
Enrico Fermi Institute and Department of Physics, University of Chicago,\\
5640 S. Ellis Avenue, Chicago IL 60637, USA}

\end{center}

\vskip 2 cm

\begin{abstract}
\baselineskip=18pt

We obtain relations among boundary states in bosonic 
minimal open string theory using the boundary ground ring. We also obtain a 
difference equation that boundary correlators must satisfy. 

\end{abstract}

\end{titlepage}

\pagestyle{plain}
%\baselineskip=18pt
% Try a wider skip
\baselineskip=19pt

\section{Introduction}

Exactly solvable string theories are of intrinsic interest,
and have also proven useful in elucidating
the structure of string theory in general. 
The basic idea is to consider two-dimensional gravity coupled to 
$c_M \leq 1$ matter conformal field theory. It is well-known that
the physical states of this theory carry non-trivial ghost 
numbers~\cite{Lian:1991gk}, and correspond to operators 
which are non-trivial in the BRST 
cohomology of the theory. Among these operators, those
with ghost number zero and conformal dimension zero form the ground ring of 
the theory, where ring multiplication corresponds to OPE in the field 
theory~\cite{Witten:1991zd}, modulo BRST exact operators. Both the 
closed and the open string sectors contain ground ring operators 
in their spectrum. In this work, we shall analyze the role of the
boundary ground ring in minimal (bosonic) string theory.  

\subsection{Summary of minimal string theory}

To begin, we briefly review certain facts about minimal string theory
which will be useful later. The world-sheet theory of two-dimensional 
gravity is given by Liouville conformal field 
theory~\cite{Curtright:1982gt}, which we couple
to ``matter'' CFT to define minimal string theory. 
Consider Liouville theory with background charge $Q_L
=b + 1/b$, where $b$ is the coupling. In addition to the kinetic term
and the curvature coupling specifying the background charge, 
the other terms in the 
action are
\be \mu \int_{M} d^2 x {\sqrt{g}} e^{2b\phi} + \mu_B \int_{\p M} dy g^{1/4}
e^{b\phi}, \ee
where $\mu$ and $\mu_B$ are the bulk and boundary cosmological constants 
respectively. The first term corresponds to the bulk Liouville interaction,
while the second yields the boundary interaction. The central charge of 
this CFT is given by 
\be c_L =1+ 6Q_L^2. \ee
Since we shall be considering boundary operators in this theory, we 
need to know their conformal dimensions. The boundary operator 
$e^{\beta \phi}$ has conformal dimension $\beta(Q_L -\beta)$. In particular, 
$\Delta (e^{b\phi}) =1$ and so the boundary interaction is a marginal
deformation. 
We want to couple this theory to matter CFT with background
charge $Q_M$. The central charge of the matter CFT is  
\be c_M =1- 6Q_M^2. \ee 
On coupling Liouville theory to 
matter, we require that $c_L +c_M =26$, which leads to $Q_M^2 =(b-1/b)^2$.
In particular, for the matter CFT we shall consider the $(p,q)$ minimal 
models~\cite{Belavin:1984vu} where $p$ and $q$ are 
relatively prime integers, and 
$p>q$. Thus we get that $b-1/b =\pm(\sqrt{p/q} -\sqrt{q/p})$.
Since $b$ is the coupling we take it to be positive, so that the theory
is free as $\phi \rightarrow -\infty$. This tells us that
\be b =\sqrt{\frac{q}{p}}.\ee
Then the boundary cosmological constant term $e^{b\phi}$ satisfies the 
Seiberg bound 
$b<Q_L/2$~\cite{Seiberg:1990eb,Polchinski:1990mh}.\footnote{The other 
choice $b=\sqrt{p/q}$ violates the bound but is a 
distinct, allowed operator \cite{Kostov:2002uq}.}

In the matter CFT, the boundary vertex operators are given by
$e^{i\alpha_{r,s} X}$ 
where
\be \label{nullmat}
\alpha_{r,s} =\frac{1}{2} (1-r) \alpha_+ + \frac{1}{2} (1-s) \alpha_-,
\ee
and $\alpha_+ =\sqrt{p/q}, \alpha_- =-\sqrt{q/p}$.
They have conformal dimension
\be \Delta (e^{i\alpha_{r,s} X}) =\alpha_{r,s} (\alpha_{r,s} -Q_M).\ee 
There are $(p-1)(q-1)/2$ such vertex operators having
$1 \leq r< q$, $1 \leq s < p$ and $pr -qs > 0$, which fill out
the Kac table of degenerate operators 
({\it c.f.} \cite{DiFrancesco:1997nk}). Using
\be Q_M =\frac{1}{b} -b = \sqrt{\frac{p}{q}} -\sqrt{\frac{q}{p}},\ee
we see that
\be \Delta (e^{i\alpha_{r,s} X}) =\frac{(pr-qs)^2 -(p-q)^2}{4pq},\ee   
which is the Kac formula.

Open string boundary conditions for the matter
also come in one-to-one correspondence
with the Kac table \cite{Cardy:1989ir}, and are thus labelled
by integers $(m,n)$, $1\leq m<q$, $1\leq n<p$.
In the Liouville field theory, the parameter $\mu_B$ enters
the boundary condition through the boundary terms
in the variation of the action.
It proves more convenient to express 
the boundary parameter in terms of the uniformization parameter $\s$ 
defined by~\cite{Fateev:2000ik}
\be \label{unpara}
\frac{\mu_B^2}{\mu} {\rm sin} \pi b^2= {\rm cosh}^2 \pi b\s.\ee  
Boundary states describe D-brane boundary conditions from the
perspective of the dual closed string channel, and are also
labelled by $(\sigma;m,n)$. 

Unlike the closed string vertex operators, in order to specify open 
string vertex operators we also need to specify the boundary conditions
across their insertion points on the boundary. So we need to specify the 
values of $\mu_B$ in the Liouville sector, and $(m,n)$ (where $1 \leq m < q$ 
and $1 \leq n < p$) in the matter sector, on either side of the insertion 
of the vertex operator on the boundary. 
So we shall express boundary vertex operators as $_{I_L}^{\s_L}[ 
{\cal{O}} ]^{\s_R}_{I_R}$,
where $\s_L$ ($\s_R$) are the values of $\s$ on the left (right) of the 
insertion, and $I_L$ ($I_R$)\footnote{$I_i \equiv (m_i,n_i)$} are the values
of $(m,n)$ on the left (right) of the insertion. Let us now mention the
constraints on the choice of these boundary conditions, which depend
on the structure of degenerate representations of the two CFTs. 

In the matter sector, if the boundary vertex operator
contains a degenerate primary of the Virasoro algebra with Kac 
label $(r,s)$, then the constraint is given by 
\be (r,s) \times (m_L,n_L) \sim (m_R,n_R),\ee
or in other words, $(m_R,n_R)$ should be in the minimal model fusion 
rule of $(r,s)$ and $(m_L,n_L)$~\cite{Cardy:1986gw,Cardy:1989ir}. 

In the Liouville sector, in order to know if there is any relation between
$\s_L$ and $\s_R$, we need to know the structure of the degenerate
representations of the Liouville Virasoro algebra. From the expression for
the Kac determinant, it follows that the degenerate representations of the
Liouville Virasoro algebra have conformal primaries $e^{\gamma_{r,s} 
\phi}$ where
\be \label{notnull} \gamma_{r,s} =\frac{(1-r)p +(1-s)q}{2\sqrt{pq}},\ee       
where $r,s \geq 1$. 
%Note that \C{tachdress} implies that
%\be \beta_{r,s} = \frac{(1-\eta r)p +(1+\eta s)q}{2\sqrt{pq}},\ee 
%where $\eta ={\rm sign} (pr-qs)$. So clearly the open string tachyons never
%contain degenerate Virasoro primaries of the Liouville CFT. 

In boundary Liouville theory, vertex operators containing degenerate 
Liouville primaries satisfying \C{notnull} have constraints 
between $\s_L$ and 
$\s_R$~\cite{Fateev:2000ik,Ponsot:2001ng}. For example, 
for the degenerate primary $e^{-b\phi/2}$ the 
restriction is that $\s_L -\s_R =\pm ib$ or $\s_L +\s_R =\pm ib$. (It 
follows that the condition for $e^{-\phi/2b}$ to be a degenerate primary
is that $\s_L -\s_R =\pm i/b$ or $\s_L +\s_R =\pm i/b$.) Thus $\s$ changes 
along the boundary across the insertion point. 
However, at the next level, for $e^{-b\phi}$ to be a degenerate primary, 
one of the possible choices is $\s_L =\s_R$ (this is also true for 
$e^{-\phi/b}$). So in this case $\s$ does not change along the boundary. 
In general for the degenerate primaries $e^{-nb\phi}$ and $e^{-n\phi/b}$, 
where $n$ is a non-negative integer, $\s_L =\s_R$ is an allowed choice.

The open string tachyon and ground ring vertex operators will
be relevant to our discussion. The open string tachyon 
vertex operators are given by
\be \label{optach} T_{r,s} =c e^{i\alpha_{r,s} X} e^{\beta_{r,s} 
\phi},\ee
where we have dropped the indices $\s_L,\s_R,I_L$ and $I_R$ for brevity.
Demanding $\Delta (T_{r,s}) =0$, we get that
\be \label{tachdress} 
\beta_{r,s} =\frac{p(1-r) +q(1+s)}{2\sqrt{pq}}.\ee
In what follows we will not be considering the other Liouville
dressing (the one that violates the Seiberg bound).
%The other sign in solving the quadratic equation for $\beta_{r,s}$ is 
%neglected as it violates the Seiberg bound.

\subsection{The boundary ground ring}

Let us next consider the elements of the boundary ground 
ring~\cite{Bershadsky:1992ub}. A generic 
element of the ground ring is given by
$^{\s_L}_{I_L}[ {\cal{O}}_{r,s} ]^{\s_R}_{I_R}$, where 
$1 \leq r < q$ and $1 \leq s < p$ and so there
are $(p-1)(q-1)$ ring elements. To construct $^{\s_L}_{I_L}[ {\cal{O}}_{r,s} 
]^{\s_R}_{I_R}$, we first consider 
$^{\s_L}_{I_L}[ V^{(L)}_{r,s} V^{(M)}_{r,s}]^{\s_R}_{I_R}$, 
where $^{\s_L}[V^{(L)}_{r,s}]^{\s_R}$ and $_{I_L}[V^{(M)}_{r,s}]_{I_R}$ are
Liouville and matter degenerate primaries respectively. 
More explicitly,
\be V^{(L)}_{r,s} = e^{\gamma_{r,s} \phi},\ee
where $\gamma_{r,s}$ is given by \C{notnull}, and 
\be V^{(M)}_{r,s} =e^{i\alpha_{r,s} X}, \ee  
where $\alpha_{r,s}$ is given by \C{nullmat}. Now it is easy
to check that $\Delta (V^{(L)}_{r,s} 
V^{(M)}_{r,s}) =1-rs$. So $^{\s_L}_{I_L}[ {\cal{O}}_{r,s} ]^{\s_R}_{I_R}$ 
is obtained
by acting on $^{\s_L}_{I_L}[ V^{(L)}_{r,s} V^{(M)}_{r,s}]^{\s_R}_{I_R}$ 
with ${\cal{L}}_{r,s}$ having conformal dimension $rs-1$ and ghost 
number zero. ${\cal{L}}_{r,s}$ is constructed entirely out of the 
$bc$ ghosts, and polynomials in $L_{-n}$ ($n \geq 1$) for both 
Liouville and matter. 
%
%Bulk ground ring operators have an analogous structure 
%(the factor ${\cal{L}}_{r,s}$
%is duplicated on both holomorphic and anti-holomorphic sides).
%The bulk ground ring is generated by the corresponding 
%${\cal{O}}_{1,2}$ and ${\cal{O}}_{2,1}$, and obeys a set of relations
%\be
%U_{q-1}({\cal{O}}_{1,2})=0\ ,\quad U_{p-1}({\cal{O}}_{2,1})=0\ ,
%\label{grrels}
%\ee
%where $U_r$ are the Chebyshev polynomials of the second kind:
%$U_{r-1}(\cos\theta)=\frac{\sin(r\theta)}{\sin\theta}$.
The boundary ground ring is generated by the
boundary ground ring elements
${\cal{O}}_{1,2}$ and ${\cal{O}}_{2,1}$, 
since according to the fusion rules
these contain all other ground ring elements in their products.
%since the matter boundary operators obey the same fusion rules.
We will not address the issue of the ring structure here,
apart from some comments in the final section.

The bulk tachyon operators form a module under the action
of the ground ring; similarly the boundary tachyons constitute
a module under the action of the boundary ground ring.
Furthermore, the boundary ground ring contains elements
which change the boundary conditions in both matter and Liouville;
thus the boundary conditions themselves are acted on
by the boundary ground ring, and we can use this fact
to deduce relations among various boundary conditions.

\subsection{The rest of the paper}

In this work, we study the role of the boundary ground ring in minimal 
string theory. To begin, in section 2 we argue that the 
disc one-point function with only one boundary insertion and no bulk 
insertions vanishes unless the boundary
vertex operator is the identity (and the
correlator is the partition function). 
In particular, this holds for 
the case when the one-point function is any 
boundary ground ring operator. 
This auxilliary result is a necessary ingredient for
the subsequent manipulations in section 3.
There, we compute disc partition functions with two boundary
ground ring insertions and no bulk insertions in two different ways, 
yielding relations among partition functions, which we postulate to lead to
an exact relation (at the level of BRST cohomology)
among boundary states in the full theory: 
\be 
\vert \s;j,k \rangle + \vert \s;j,k-2 \rangle=
\vert \s +ib;j,k-1 \rangle + \vert \s -ib;j,k-1 \rangle, \qquad k
\geq 2. \ee 
The relation among boundary states proposed 
in~\cite{Seiberg:2003nm} follows by linear superposition
from this equality together with an analogous one 
involving shifts of the $j$ index and shifts of $\sigma$
by $i/b$. 

In section 4, we construct a difference equation 
that is satisfied by a certain class of correlators in minimal string 
theory, again by evaluating the ground ring operators'
action on the tachyons in the correlator in different ways. 
The most general correlators can be analyzed along similar 
lines. This generalizes a similar equation obtained 
in~\cite{Kostov:2003cy}, where a Neumann boundary condition was considered 
for the matter CFT. Finally we end with a discussion.

%%%%%%%%%%%%%%%%%%%%%%%%%%%%%%%%%%%%%%%%%%%%%%%%%%%%%%%%%%%%%%%%%%
%%%%%%%%%%%%%%%%%%%%%%%%%%%%%%%%%%%%%%%%%%%%%%%%%%%%%%%%%%%%%%%%%%
         
\section{Disc correlators with only one boundary insertion}

In this section, we discuss disc correlators with no closed string 
vertex operators and with only one open string vertex operator inserted on 
the boundary. 
There are particular ground ring elements for which 
$\s$ and $(m,n)$ do not change across their insertion points. 
From our previous discussion, we see that they correspond to the 
choices $r=2l+1,s=2k+1$, where $0\leq 2l <q-1$ and $0\leq 2k <p-1$.
Hence the disc one-point 
function for such an operator does not naively vanish by any symmetry 
arguments. In particular, considering the boundary ground ring
elements, we see that the choice 
$k=l=0$ gives the disc partition function with $\s$ and $(m,n)$ 
boundary conditions, which is of course non-vanishing. We now show that 
\be \label{ringvan}
\langle _{(m,n)}^\s[ {\cal{O}}_{2l+1,2k+1} 
]^\s_{(m,n)} \rangle = 0,\ee
for $k,l \geq 1$, which will be useful later. Also it will follow trivially
from our analysis that this statement is true for any such one-point 
function: except the partition function, all disc amplitudes with
only one boundary insertion vanish. 
 
In order to compute the disc correlator for the ground ring, 
we consider the equation
\be \label{bdycor}
\frac{\p^2}{\p \mu_B^2} \langle ^\s_{(m,n)}[ {\cal{O}}_{2l+1,2k+1} 
]^\s_{(m,n)} 
\rangle \vert_\mu = \langle ^\s_{(m,n)}[ T_{1,1} ]^\s_{(m,n)}
[ T_{1,1} ]^\s_{(m,n)} [ {\cal{O}}_{2l+1,2k+1} ]^\s_{(m,n)} \rangle , \ee 
where $^\s_{(m,n)}[ T_{1,1} ]^\s_{(m,n)}
= c~ ^\s_{(m,n)}[ e^{b\phi} ]^\s_{(m,n)}$ is the boundary 
Liouville interaction. 
We have inserted two boundary tachyons, which fixes a part of 
the conformal killing group of the disc leaving a residual finite
part, so the correlator does not vanish trivially. 
The three-point function 
is given by ${\cal{L}}_{2l+1,2k+1}$ acting on a product of Liouville and matter
three-point functions. 
The Liouville correlator can be computed using the fact that the 
ground ring element 
is a degenerate primary (this approach was originally proposed by Teschner 
in computing bulk Liouville correlators~\cite{Teschner:1995yf}) and
yields a finite number of terms in its OPE with $T_{1,1}$, one of
which is $T_{1,1}$ itself. Thus the correlator reduces to a 
two-point function of $T_{1,1}$'s which is 
non-vanishing~\cite{Fateev:2000ik}. One can also calculate it directly 
using the expression for the  
boundary three-point function~\cite{Ponsot:2001ng}. 

Another way to do the computation is to consider the equation
\be \label{bulkint}
\frac{\p}{\p \mu} \langle ^\s_{(m,n)}[ {\cal{O}}_{2l+1,2k+1} 
]^\s_{(m,n)} \rangle 
\vert_{\mu_B} = -\langle (c \bar{c} e^{2b\phi}) ^\s_{(m,n)}[ 
{\cal{O}}_{2l+1,2k+1} ]^\s_{(m,n)} \rangle , \ee 
and compute the bulk-boundary structure constant using expressions
given in~\cite{Hosomichi:2001xc,Ponsot:2003ss}. So the 
Liouville part of the one-point function is non-vanishing and finite.

In order to compute the matter correlator, let us consider the
three-point function of the Virasoro primaries 
\be \label{mthree}
\langle ^A[ V^{\sst {(M)}}_I ]^B [ V^{\sst {(M)}}_J ]^C 
[ V^{\sst {(M)}}_K]^A \rangle =C_{IJK}^{ABC},\ee
for any rational CFT, where we are interested in the particular
case when two of the operators are 
identity. Alternatively, we could have also
considered the bulk-boundary correlator
\be \label{mbbthree}
\langle {\cal{V}}^{\sst {(M)}}_I ~^A[ V^{\sst {(M)}}_J ]^A \rangle, \ee  
where ${\cal{V}}^{\sst {(M)}}$ is a bulk operator, 
which is the identity in our case.

The structure constant \C{mthree} can be deduced and essentially depends 
on the fusion (F) matrix of the RCFT.\footnote{The 
bulk-boundary structure 
constant \C{mbbthree} depends on the fusion matrix and also on the modular
matrix element for the torus with one operator insertion.} In fact, using
Lewellen's sewing constraints~\cite{Lewellen:1991tb} on the 
disc, the structure constant \C{mthree} has been deduced 
in~\cite{Runkel:1998pm} for the 
A--series Virasoro minimal models (the generalization to the D--series has 
also been done in~\cite{Runkel:1999dz}). Upto freedom of
rescaling fields, Runkel has argued that
\be \label{mm6j}
C_{IJK}^{ABC} =F_{BK} 
\left[ \begin{array}{cc}
A & C \\
I & J 
\end{array}\right] .
\ee 
The same conclusion has also been reached by other 
authors~\cite{Behrend:1999bn,Felder:1999mq}, who
have used different normalizations. 
In~\cite{Behrend:1999bn}, the ratio of the three-point function to 
the two-point function is given by the fusion matrix
\be 
\frac{C_{IJK}^{ABC}}{C_{KK}^{AC}} =F_{BK} 
\left[ \begin{array}{cc}
A & C \\
I & J 
\end{array}\right] , \ee
while in~\cite{Felder:1999mq}, the authors have normalized the 
correlator as
\be 
\frac{C_{IJK}^{ABC}}{C_{IJK}} = 
\sqrt{\frac{S_{I0} S_{J0}}{S_{K0} S_{00}}} \left\{ \begin{array}{lll}
I & A & B \\
C & J & K \\
\end{array} \right\} ={\sqrt{\frac{S_{I0} S_{J0}}{S_{K0} S_{00}}}}
F_{BK} 
\left[ \begin{array}{cc}
A & C \\
I & J 
\end{array}\right] ,\ee 
where $S_{IJ}$ is the modular matrix, $\{ \cdots \}$ is the 
quantum $6j$ symbol and $C_{IJK}$ is the structure constant for the bulk
three-point function on the sphere. The last equality in the above equation 
follows from the discussion below. 

Using the quantum group approach to studying rational CFTs, it can be 
shown that the quantum $6j$ symbols solve the polynomial equations of 
Moore and Seiberg~\cite{Moore:1988qv,Moore:1989vd}. In fact, 
the fusion matrix is related to
the quantum $6j$ symbol by the 
equation~\cite{Alvarez-Gaume:1988vr,Alvarez-Gaume:1989aq}
\be 
F_{IJ} 
\left[ \begin{array}{cc}
K & L \\
M & N 
\end{array}\right] =\left\{ \begin{array}{lll}
M & K & I \\
L & N & J \\
\end{array} \right\}. \ee
The definition of the $6j$ symbol (see~\cite{abook} for its explicit 
form) involves a parameter $q$ of the
quantum group $SU_q (2)$.\footnote{Not to be confused
with the label $q$ of the $(p,q)$ minimal model.} 

%The $(p,q)$ minimal models have also 
%been analyzed using the quantum group 
%approach~\cite{Gomez:1990er,Gomez:1989sw}. These 
%theories have a quantum group symmetry, which is formulated in terms 
%of screened vertex operators.\footnote{These vertex operators are different 
%from those constructed by Felder~\cite{Felder:1988zp} in the choice 
%of contours} The thermal operators $\alpha_{1,n}$ ($\alpha_{n,1}$) 
%form a representation of the $SU_{q_+} (2)$ $ (SU_{q_-} (2))$ subgroup 
%of the full quantum group, with $q_+ =e^{2\pi i \alpha_+^2}$ and 
%$q_- =e^{2\pi i \alpha_-^2}$. Also, the fusion matrices factorize into 
%a product of two quantum $6j$ symbols~\cite{Furlan:1989ra,Furlan:1990dh}.

So for our purposes, from \C{bdycor} the three-point function is
proportional to
\be 
C_{(1,1)(1,1)(2l+1,2k+1)}^{(m,n)(m,n)(m,n)} =F_{(m,n)(2l+1,2k+1)} 
\left[ \begin{array}{cc}
(1,1) & (1,1) \\
(m,n) & (m,n) 
\end{array}\right] =0.\ee 
This vanishes for $k,l \geq 1$ trivially as $(1,1) \times (1,1) \rightarrow
(2l+1,2k+1)$ has vanishing fusion rule, leading to \C{ringvan}. In fact, from 
\C{mm6j} it follows that the boundary three-point function is
non-vanishing only if the three boundary indices $I,J$ and $K$ have non-zero 
fusion rule. So all boundary one-point functions vanish (except the
partition function where the three operators are all the identity) on
the disc.    

\section{Basic relations among boundary states}

We now use the boundary ground ring to motivate various relations 
among boundary states, for which we also present some evidence. The 
general strategy is to consider the disc amplitude with arbitrary bulk vertex
operator insertions, and two boundary ground ring insertions compatible
with the choice of boundary conditions. Using 
the fact that any correlator is independent of the position
of insertion of a ground ring element, we can contract the two ground ring 
elements in two ways along the boundary, see figure \twoways. 

%%%%%%%%%%%%%%%%%%%%%%%%%%%%%%%%%%%%%%%%%%%%%%%%%%%%%%%5
\begin{figure}[ht]
%\label{twoways}
\begin{center}
\[
\mbox{\begin{picture}(270,125)(0,0)
\includegraphics[scale=.5]{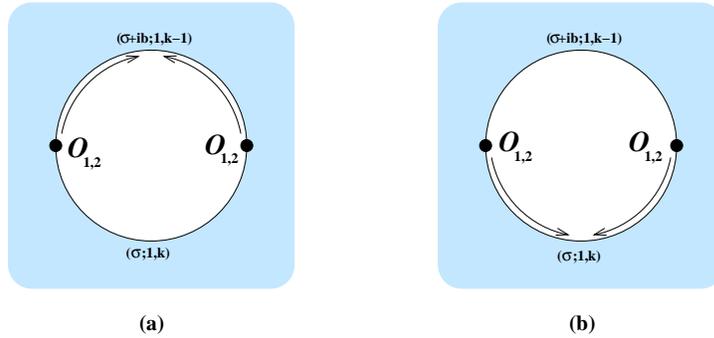}
\end{picture}}
\]
\caption{\it
Utilizing two ways of contracting ground ring operators along the boundary
yields an identity on correlation functions.
}
\end{center}
\end{figure}
%%%%%%%%%%%%%%%%%%%%%%%%%%%%%%%%%%%%%%%%%%%%%%%%%%%%%%%

This gives a relation
among correlators with the same bulk insertions but different boundary 
insertions and different boundary conditions. We shall obtain such 
relations involving partition functions,
and motivate relations among boundary states in the full theory from 
them. 

To begin with, we consider the two-point functions
\be \label{twoptdif}
\langle ^{\s +ib}_{(1,1)}[ {\cal{O}}_{1,2} ]^\s_{(1,2)}
[ {\cal{O}}_{1,2} ]^{\s +ib}_{(1,1)} \rangle, \quad
\langle ^{\s -ib}_{(1,1)}[ {\cal{O}}_{1,2} ]^\s_{(1,2)}
[ {\cal{O}}_{1,2} ]^{\s -ib}_{(1,1)} \rangle , \ee 
and evaluate them separately. Here on one segment of the disc we have 
boundary conditions $\{\s, (1,2)\}$, while on the other segment we consider
both the possibilities $\s \pm ib$ allowed by the insertion of 
${\cal{O}}_{1,2}$, while fixing the matter boundary condition to be 
$(1,1)$. This construction will motivate a certain relation among 
correlators, leading us to generalize this analysis later.   

Each correlator factorizes into a product
of Liouville and matter correlators. We shall neglect all factors that are
independent of the boundary conditions, so factors arising from
${\cal{L}}_{1,2}$ will be dropped. We first evaluate
\be \langle ^{\s +ib}_{(1,1)}[ {\cal{O}}_{1,2} ]^\s_{(1,2)}
[ {\cal{O}}_{1,2} ]^{\s +ib}_{(1,1)} \rangle, \ee
by contracting ${\cal{O}}_{1,2} .{\cal{O}}_{1,2}$ along $(\s +ib;1,1)$.
This yields\footnote{In all these calculations involving no 
bulk operator insertions, only the
partition function survives the contraction-all the other one-point 
functions involve the boundary operator ${\cal{O}}_{1,3}$ and 
vanish using the arguments of the previous section.} 

\be 
\langle ^{\s +ib}_{(1,1)}[ {\cal{O}}_{1,2} ]^\s_{(1,2)}
[ {\cal{O}}_{1,2} ]^{\s +ib}_{(1,1)} \rangle = \lambda_L \lambda_M
Z (\s;1,2),\ee
where 
\bea \label{pertcal}
&&^\s[ e^{-b\phi/2} ]^{\s +ib} [ e^{-b\phi/2} ]^\s \sim 
{^\s}[ e^{-b\phi} ]^\s +\lambda_L ~ ^\s[1]^\s , \non \\
&&_{(1,2)}[ e^{ibX/2} ]_{(1,1)} [ e^{ibX/2} ]_{(1,2)} \sim 
{_{(1,2)}}[ e^{ibX} ]_{(1,2)} +\lambda_M ~ _{(1,2)}[1]_{(1,2)}. \eea

On the other hand, contracting ${\cal{O}}_{1,2} .{\cal{O}}_{1,2}$
along $(\s;1,2)$ yields
\be \langle ^{\s +ib}_{(1,1)}[ {\cal{O}}_{1,2} ]^\s_{(1,2)}
[ {\cal{O}}_{1,2} ]^{\s +ib}_{(1,1)} \rangle = \lambda_L' \lambda_M'
Z (\s +ib;1,1),\ee
where
\bea &&^{\s +ib}[ e^{-b\phi/2} ]^\s [ e^{-b\phi/2} ]^{\s +ib} \sim 
{^{\s +ib}}[ e^{-b\phi} ]^{\s +ib} +\lambda_L' ~ ^{\s +ib}[1]^{\s +ib} 
, \non \\ &&_{(1,1)}[ e^{ibX/2} ]_{(1,2)} [ e^{ibX/2} ]_{(1,1)} \sim 
{_{(1,1)}}[ e^{ibX} ]_{(1,1)} +\lambda_M' ~ _{(1,1)}[1]_{(1,1)}, \eea
leading to the relation
\be \label{vevone}
\langle ^{\s +ib}_{(1,1)}[ {\cal{O}}_{1,2} ]^\s_{(1,2)}
[ {\cal{O}}_{1,2} ]^{\s +ib}_{(1,1)} \rangle = \lambda_L 
\lambda_M Z (\s;1,2) = \lambda_L' \lambda_M'
Z (\s +ib;1,1) .\ee
Similarly for the other correlator in \C{twoptdif}, we get
\be \label{vevtwo}
\langle ^{\s -ib}_{(1,1)}[ {\cal{O}}_{1,2} ]^\s_{(1,2)}
[ {\cal{O}}_{1,2} ]^{\s -ib}_{(1,1)} \rangle = {\widetilde{\lambda}}_L 
\lambda_M Z (\s;1,2) = {\widetilde{\lambda}}_L' \lambda_M'
Z (\s -ib;1,1) ,\ee
where
\bea &&^\s[ e^{-b\phi/2} ]^{\s -ib} [ e^{-b\phi/2} ]^\s \sim 
{^\s}[ e^{-b\phi} ]^\s +{\widetilde{\lambda}}_L ~ ^\s[1]^\s , \non \\
&&^{\s-ib}[ e^{-b\phi/2} ]^\s [ e^{-b\phi/2} ]^{\s -ib} \sim 
{^{\s -ib}}[ e^{-b\phi} ]^{\s -ib} +{\widetilde{\lambda}}_L' 
~ ^{\s -ib}[1]^{\s -ib}.\eea
Thus from \C{vevone} and \C{vevtwo} we get 
\be \label{relpart}
Z (\s +ib;1,1) + Z (\s -ib;1,1) = \frac{\lambda_M}{\lambda_M'}
\left( \frac{\lambda_L}{\lambda_L'} 
+\frac{{\widetilde\lambda}_L}{{\widetilde\lambda}_L'} 
\right) Z (\s ;1,2).\ee

We now proceed to calculate the various coefficients in this relation.
From \C{pertcal}, we see that $\lambda_L$ can be calculated 
perturbatively using screening 
integrals~\cite{Feigin:1981st,Dotsenko:1984nm,Dotsenko:1984ad}. We get
\be \label{calcl}
-\lambda_L = \mu_B^i \int_{C_i} dx \langle ^\s[ e^{-b\phi/2} 
(0) ]^{\s +ib} [ e^{-b\phi/2} (1) ]^\s [e^{Q_L \phi} (\infty)]^\s 
[e^{b\phi} (x)]\rangle, \ee
where
\be
\mu_B^i = \left\{ \begin{array}{ll}
{\sqrt \mu}  {\rm cosh} \pi b \s, & C_i \in (-\infty,0),\\
{\sqrt \mu}  {\rm cosh} \pi b (\s +ib), & C_i \in [0,1),\\
{\sqrt \mu}  {\rm cosh} \pi b \s, & C_i \in [1,\infty).
\end{array} \right. \ee

Now \C{calcl} can be evaluated using free field theory, and gives
\be \lambda_L =-{\sqrt \mu} B(1+b^2,1+b^2)\left[ {\rm cosh} 
\pi b (\s +ib) -\frac{{\rm cosh} \pi b \s}{{\rm cos}\pi b^2}\right].\ee

Similarly we get
\bea \lambda_L' =-{\sqrt \mu} B(1+b^2,1+b^2)\left[ {\rm cosh} 
\pi b\s -\frac{{\rm cosh} \pi b (\s +ib)}{{\rm cos}\pi b^2}\right], \non \\
{\widetilde\lambda}_L =-{\sqrt \mu} B(1+b^2,1+b^2)\left[ {\rm cosh} 
\pi b (\s -ib) -\frac{{\rm cosh} \pi b \s}{{\rm cos}\pi b^2}\right], \non \\
{\widetilde\lambda}_L' =-{\sqrt \mu} B(1+b^2,1+b^2)\left[ {\rm cosh} 
\pi b\s -\frac{{\rm cosh} \pi b (\s -ib)}{{\rm cos}\pi b^2}\right].
\eea

Straightforward algebra shows that
\be \frac{\lambda_L}{\lambda_L'} 
+\frac{{\widetilde\lambda}_L}{{\widetilde\lambda}_L'} =-2 {\rm cos} \pi 
b^2.\ee
Note that this combination is independent of $\s$. For $\lambda_M$
and $\lambda_M'$, using the normalization in \C{mm6j}, we get
\be \lambda_M =F_{(1,1)(1,1)} 
\left[ \begin{array}{cc}
(1,2) & (1,2) \\
(1,2) & (1,2) 
\end{array}\right], \ee
and
\be \lambda_M' = F_{(1,2)(1,1)} 
\left[ \begin{array}{cc}
(1,1) & (1,1) \\
(1,2) & (1,2) 
\end{array}\right]. \ee
In any RCFT, we have the relations~\cite{Moore:1989vd}
\be \label{rcftf0} F_{00} 
\left[ \begin{array}{cc}
I & I \\
I & I 
\end{array}\right] =\frac{S_{00}}{S_{0I}},\ee
and
\be \label{rcftf} F_{K0} 
\left[ \begin{array}{cc}
I & I \\
J & J 
\end{array}\right] =\frac{S_{0K}}{S_{0J}} 
F_{J0} \left[ \begin{array}{cc}
K & K \\
I & I 
\end{array}\right],\ee
leading to
\be \frac{\lambda_M}{\lambda_M'} 
=\frac{S_{(1,1)(1,1)}}{S_{(1,1)(1,2)}}.\ee
Then \C{relpart} gives
\be \label{difconv}
Z (\s +ib;1,1) +Z (\s -ib;1,1) =-2 {\rm cos} \pi 
b^2 \frac{S_{(1,1)(1,1)}}{S_{(1,1)(1,2)}}
Z(\s;1,2).\ee
Using the expression for the modular matrix
\be \label{modmat}
S_{(r,s)(r',s')} =2\sqrt{\frac{2}{pq}} (-1)^{1+rs' +r's} {\rm sin} 
\frac{\pi prr'}{q} {\rm sin} \frac{\pi qss'}{p}, \ee
we see that
\be \frac{S_{(1,1)(1,1)}}{S_{(1,1)(1,2)}} =-\frac{1}{2 {\rm cos} \pi 
b^2}.  \ee
Thus \C{difconv} leads to
\be \label{withvev}
\langle 0 \vert \s;1,2 \rangle =
\langle 0 \vert \s +ib;1,1 \rangle + \langle 0 \vert \s -ib;1,1\rangle.\ee
We have thus arrived at a relation 
between correlators with different Liouville
and matter content, where the shift in matter content is accompanied
by an imaginary shift in $\s$. We now obtain a generalization of
\C{withvev} to all disk partition functions.

We consider the correlator
\be \langle ^{\s +ib}_{(1,k-1)}[ {\cal{O}}_{1,2} ]^\s_{(1,k)}
[ {\cal{O}}_{1,2} ]^{\s +ib}_{(1,k-1)} \rangle, \ee
and evaluate it in two ways as mentioned above. We get that
\be 
\langle ^{\s +ib}_{(1,k-1)}[ {\cal{O}}_{1,2} ]^\s_{(1,k)}
[ {\cal{O}}_{1,2} ]^{\s +ib}_{(1,k-1)} \rangle = \lambda_L 
{\widetilde\lambda}_M Z (\s;1,k) = \lambda_L' {\widetilde\lambda}_M'
Z (\s +ib;1,k-1) ,\ee
where
\be {\widetilde\lambda}_M =F_{(1,k-1)(1,1)} 
\left[ \begin{array}{cc}
(1,k) & (1,k) \\
(1,2) & (1,2) 
\end{array}\right], \ee
and
\be {\widetilde\lambda}_M' = F_{(1,k)(1,1)} 
\left[ \begin{array}{cc}
(1,k-1) & (1,k-1) \\
(1,2) & (1,2) 
\end{array}\right]. \ee
Thus using \C{rcftf}, we get
\be \frac{{\widetilde\lambda}_M}{{\widetilde\lambda}_M'}
=\frac{S_{(1,1)(1,k-1)}}{S_{(1,1)(1,k)}}. \ee

We also evaluate the correlator with two boundary ${\cal{O}}_{1,2}$ 
insertions,
$(\s,\s -ib)$ boundary conditions and same matter boundary conditions, to 
get
\be 
\langle ^{\s -ib}_{(1,k-1)}[ {\cal{O}}_{1,2} ]^\s_{(1,k)}
[ {\cal{O}}_{1,2} ]^{\s -ib}_{(1,k-1)} \rangle = {\widetilde\lambda}_L 
{\widetilde\lambda}_M Z (\s;1,k) = {\widetilde\lambda}_L' 
{\widetilde\lambda}_M' Z (\s -ib;1,k-1) .\ee

Thus
\be \label{twoOk}
Z (\s +ib;1,k-1) + Z (\s -ib;1,k-1) = -2 {\rm cos} \pi b^2
\frac{S_{(1,1)(k-1)}}{S_{(1,1)(1,k)}} Z (\s ;1,k).\ee

Unlike \C{withvev}, the right-hand side of \C{twoOk} 
depends on the boundary condition in the matter sector. We would like
to deduce an equation free from this constraint, and so using
the relation in the matter sector
\be Z (m,n) =\frac{S_{(1,1)(m,n)}}{S_{(1,1)(m',n')}} Z (m',n'), \ee 
which follows from~\cite{Cardy:1989ir}
\be Z (m,n) =\frac{S_{(1,1)(m,n)}}{\sqrt{S_{(1,1)(1,1)}}},\ee
we rewrite \C{twoOk} as
\bea \label{vevmore}
Z (\s +ib;1,k-1) + Z (\s -ib;1,k-1) &=& -2 {\rm cos} \pi b^2
\frac{S_{(1,1)(1,k-1)}}{S_{(1,1)(1,k)}} Z (\s ;1,k) \non \\
&=& -2 {\rm cos} \pi b^2
\frac{S_{(1,1)(1,k-1)}}{S_{(1,1)(1,k-2)}} Z (\s ;1,k-2).
\eea

From \C{vevmore}, it follows that
\bea \label{relZs}
Z (\s ;1,k) + Z (\s ;1,k-2) &=& -\frac{1}{2 {\rm cos} \pi b^2}
\frac{S_{(1,1)(1,k)} +S_{(1,1)(k-2)}}{S_{(1,1)(k-1)}} \non \\ &&\times 
\left[ Z (\s +ib;1,k-1) + Z (\s -ib;1,k-1) \right]. \eea

On using \C{modmat}, we simplify \C{relZs} to get 
\be \label{relcorr} 
\langle 0 \vert \s;1,k \rangle +\langle 0 \vert \s;1,k-2 \rangle=
\langle 0 \vert \s +ib;1,k-1 \rangle + \langle 0 \vert \s -ib;1,k-1
\rangle,\ee
for $k \geq 2$ (in fact \C{withvev} is the particular case with $k=2$).
Note that this equation relates correlators with coefficients independent 
of boundary conditions, and involves only $(1,k),(1,k-1),(1,k-2)$ Cardy 
boundary states and $(\s,\s +ib,\s-ib)$ FZZT boundary states. Applied 
recursively, it gives a set of relations among correlators.

Now \C{relcorr} by itself is a relation among disc
partition functions, which we 
want to elevate to a relation among boundary states in the full theory, 
modulo BRST exact states. We propose the relation among boundary states 
\be \label{eqstate}
\vert \s;1,k \rangle + \vert \s;1,k-2 \rangle=
\vert \s +ib;1,k-1 \rangle + \vert \s -ib;1,k-1 \rangle, \qquad k 
\geq 2. \ee
We have established this equality at the level of the disc partition function 
above. To provide more evidence, we calculate the disc one-point function
with one bulk vertex operator insertion and check the equality. This 
analysis is very similar to the one in~\cite{Seiberg:2003nm}, to 
which we shall soon return.
 
The boundary state $\vert \s;1,s \rangle$ is given by
\be \vert \s;1,s \rangle = \sum_{m,n}
\int_0^\infty \frac{dp}{\pi} {\rm cos} 
(2\pi\s p) U^*(p) \frac{S_{(1,s)(m,n)}}{\sqrt{S_{(1,1)(m,n)}}}
\vert p \gg \vert m,n \gg,\ee
where $\vert p \gg$ and $\vert m,n \gg$ are Liouville and matter Ishibashi 
states~\cite{Ishibashi:1988kg,Onogi:1988qk} respectively, 
and~\cite{Fateev:2000ik}
\be U (p) =(\pi \mu \gamma (b^2))^{-ip/b}\frac{\Gamma (1+2ipb) \Gamma 
(1+\frac{2ip}{b})}{ip}.\ee 

Considering the bulk tachyon vertex operator
\be {\mathcal{T}}_{r,s} =c \bar{c} e^{2i\alpha_{r,s} X} e^{2\beta_{r,s} 
\phi},\ee

and dropping all factors independent of the boundary conditions, we get
\bea \label{explcalc}
\langle {\mathcal{T}}_{r,s} \vert \s;1,k \rangle 
+\langle {\mathcal{T}}_{r,s} \vert \s;1,k-2 \rangle=
\langle {\mathcal{T}}_{r,s} \vert \s +ib;1,k-1 \rangle 
+ \langle {\mathcal{T}}_{r,s} \vert \s -ib;1,k-1 \rangle \non \\
= (-1)^{s+rk} {\rm cosh} \frac{\pi\s(pr-qs)}{\sqrt{pq}} 
{\rm sin} 
\frac{\pi rp}{q} \left[ {\rm sin} \frac{\pi skq}{p} + {\rm sin}  
\frac{\pi s(k-2)q}{p}\right]. \eea

We can similarly verify the equality with a bulk ground ring insertion, and
the insertion of a bulk physical operator at negative ghost number -- the
answer goes through with 
appropriate changes in the range of $s$~\cite{Seiberg:2003nm}. Since these
are all the physical operators in the theory, the equality \C{explcalc}
provides evidence for \C{eqstate}. 

From \C{eqstate} it follows that a boundary state with Liouville 
and matter content $(\s;1,k)$ is given by a linear combination of 
boundary states with matter content $(1,k-1)$ and $(1,k-2)$ with 
$\s$ appropriately shifted. So using this relation recursively, we can
express a boundary state with content $(\s;1,k)$ as a sum of 
boundary states with different $\s$ but trivial matter content -- the 
$(1,1)$ state. In fact for $k=2$, we already have the expression
\be \label{k=2} \vert \s;1,2 \rangle = 
\vert \s + ib ;1,1 \rangle +\vert \s - ib ;1,1 \rangle. \ee
For $k=3$, \C{eqstate} and \C{k=2} gives
\be \label{k=3} \vert \s;1,3 \rangle = \vert \s +2ib ;1,1 \rangle
+ \vert \s  ;1,1 \rangle + \vert \s -2ib ;1,1 \rangle ,\ee     
while for $k=4$, \C{eqstate}, \C{k=2} and \C{k=3} gives
\be \vert \s;1,4 \rangle = \vert \s +3ib ;1,1 \rangle
+ \vert \s  +ib;1,1 \rangle + \vert \s  -ib;1,1 \rangle 
+ \vert \s -3ib ;1,1 \rangle,\ee 
and so on. In general, we have
\be \label{lowmat1}
\vert \s ; 1,k \rangle = \sum_m \vert \s +imb; 1,1 \rangle, \ee
where
\be \label{rangem} m = -(k-1), -(k-1) +2, \ldots, k-1. \ee
It is easy to see that \C{lowmat1} solves \C{eqstate} because
\bea 
\vert \s;1,k \rangle + \vert \s;1,k-2 \rangle=
\vert \s +ib;1,k-1 \rangle + \vert \s -ib;1,k-1 \rangle \non \\
=\vert \s -ib(k-1);1,1 \rangle + \vert \s +ib(k-1);1,1 \rangle
+2 \sum_{t} \vert \s +ibt; 1,1 \rangle, \eea 
where
\be t = -(k-3), -(k-3) +2, \ldots, k-3. \ee

Replacing ${\cal{O}}_{1,2}$ with ${\cal{O}}_{2,1}$ 
in the arguments above, gives the other relation 
\be \label{eqstate2}
\vert \s; l,1 \rangle + \vert \s; l-2,1 \rangle=
\vert \s +\frac{i}{b} ;l-1,1 \rangle + \vert \s -\frac{i}{b} ;l-1,1 
\rangle, \qquad l 
\geq 2, \ee
leading to
\be \label{lowmat2}
\vert \s ; l,1 \rangle = \sum_n \vert \s +\frac{in}{b}; 1,1 \rangle, \ee
where
\be \label{rangen}
n = -(l-1), -(l-1) +2, \ldots, l-1. \ee 

Using the independence of shifts by $b$ and $1/b$, \C{lowmat1} 
and \C{lowmat2} together give

\be \label{lowmat}
\vert \s ; l,k \rangle = \sum_{m,n} \vert \s +imb +\frac{in}{b}
; 1,1 \rangle, \ee
where $m$ and $n$ are given by \C{rangem} and \C{rangen}.
Now \C{lowmat} relates a boundary state with arbitrary Liouville and matter 
content to a sum of boundary states with Liouville content 
changed by imaginary shifts in $\s$, and trivial matter content. This 
is precisely the relation obtained by Seiberg and Shih 
in~\cite{Seiberg:2003nm}, which follows as a consequence of \C{eqstate}
and \C{eqstate2}. 

We see that the relation \C{lowmat} follows naturally using the 
boundary ground ring. Essentially in \C{lowmat} the shift in $\s$ is given 
by multiples of $b$ and $1/b$, which is the same as the change 
$\s_L -\s_R$ (or $\s_L +\s_R$) across insertions of boundary ground ring
elements.  
In fact, across the insertion of the ring
element ${\cal{O}}_{l,k}$, the change in $\s$ is given by $\s_L -\s_R$ (or
$\s_L +\s_R$) =$imb +in/b$~\cite{Fateev:2000ik}, where 
$m$ and $n$ are exactly as given in 
\C{rangem} and \C{rangen}. Thus such a shift relation has a natural 
interpretation as shifts due to the degenerate structure of the boundary 
ground ring.

\section{A difference equation for boundary correlators}

The boundary ground ring can also be used to constrain
correlation functions involving boundary vertex operator 
insertions~\cite{Bershadsky:1992ub}.  For simplicity, we 
restrict the discussion to
disc correlators which contain only boundary vertex 
operators.  An equation satisfied by such correlators has 
been obtained in~\cite{Kostov:2003cy}, where a Neumann boundary condition 
was imposed for the matter CFT, which we generalize to Liouville 
theory coupled to minimal models.\footnote{Neumann boundary conditions
are not among the Cardy boundary conditions.}
We consider certain correlators and 
show that they satisfy a difference equation using the fact 
that the tachyons form a module under the action of the ground 
ring~\cite{Kutasov:1991qx,Bershadsky:1992ub} in both open and 
closed string theory. 

The two basic ground ring elements are  
\bea \label{ringbasic}
_{(m,n)}^\s[ {\cal{O}}_{1,2}]_{(m,n \pm 1)}^{\s \pm ib}  = 
\left( bc -\frac{1}{b} \p (\phi +iX)\right) {^\s[} e^{-b\phi/2}]^{\s \pm ib} 
{_{(m,n)}[} e^{ibX/2}]_{(m,n \pm 1)} ,\non \\
^\s_{(m,n)}[{\cal{O}}_{2,1}]_{(m \pm 1,n)}^{\s \pm i/b} = 
\left( bc -b\p (\phi -iX)\right) ^\s[ e^{-\phi/2b}{]^{\s \pm i/b}}
{_{(m,n)}[} e^{-iX/2b}]_{(m \pm 1,n)}.\eea 

First let us consider the theory with $\mu_B =0$. Among the boundary 
tachyons \C{optach}, the ones given by
$T_{r,1}$ ($1 \leq r <q$) and $T_{q-1,s}$ ($1 \leq s < p-p/q$) form 
submodules of the tachyon module (this follows exactly as discussed 
in~\cite{Kutasov:1991qx} for the closed string tachyons). 
The action of the ground ring on these tachyons is given by 
\bea \label{tachmod1}
 _{(m,n)}[ {\cal{O}}_{1,2}]_{(m,n \pm 1)} 
[ T_{r,1}]_{(m',n \pm 1)} =0, \qquad \qquad \qquad \qquad \qquad \qquad
\non \\
_{(m,n)}[ {\cal{O}}_{1,2}]_{(m,n \pm 1)} [ T_{q-1,s}
]_{(m,n')} = F_{(m,n \pm 1)(q-1,s-1)} 
\left[ \begin{array}{cc}
(m,n) & (m,n') \\
(1,2) & (q-1,s) 
\end{array}\right]
{_{(m,n)}[} T_{q-1,s-1} ]_{(m,n')}. \eea

Obviously, in the first equation, $(m',n \pm 1)$ must be in the 
fusion rule of $(m,n \pm 1) \times (r,1)$, while in the second 
equation $(m,n')$ must be in the fusion rule of $(m,n \pm 1) \times 
(q-1,s)$ as well as $(m,n) \times (q-1,s-1)$. Analogous statements
hold for the equations that follow.  
Also
\bea _{(m,n)}[{\cal{O}}_{2,1}]_{(m \pm 1,n)} [ T_{q-1,s} ]_{(m \pm 1,n')} = 0,
\qquad \qquad \qquad \qquad \qquad \qquad
\non \\  _{(m,n)}[{\cal{O}}_{2,1}]_{(m \pm 1,n)} [ T_{r,1}]_{(m',n)} 
= F_{(m \pm 1,n)(r+1,1)} 
\left[ \begin{array}{cc}
(m,n) & (m',n) \\
(2,1) & (r,1) 
\end{array}\right] 
{_{(m,n)}[} T_{r+1,1}]_{(m',n)}.\eea

As before, we have dropped factors independent of boundary 
conditions -- they can be absorbed in the definition of the tachyons, or 
by suitably rescaling the fusion matrices. The coefficients in these
equations are boundary three-point functions and so are
given by the fusion matrices. Note that under the interchange of the 
boundary tachyon and ${\cal{O}}_{1,2}$ (or ${\cal{O}}_{2,1}$)
one picks up a sign\cite{Bershadsky:1992ub}. For example,   
\be _{(m,n)}[ T_{r,1}]_{(m',n)} [ {\cal{O}}_{2,1}]_{(m' \pm 1,n)} 
= - F_{(m',n)(r+1,1)} 
\left[ \begin{array}{cc}
(m,n) & (m' \pm 1,n) \\
(r,1) & (2,1) 
\end{array}\right]
{_{(m,n)}[} T_{r+1,1} ]_{(m' \pm 1,n)}.\ee

For the correlators we consider, we need the action of the ground 
ring on the tachyon module in the presence
of ordered integrated boundary vertex operators. We shall consider 
correlators with a certain ordering, namely the ones which are 
independent of the position of insertion of the ground ring 
elements.\footnote{This is not true for the others, because of
non-vanishing contributions from the boundary of moduli 
space~\cite{Bershadsky:1992ub}.} In particular, we need to know if the 
first equation in \C{tachmod1} gets modified. For the correlators
we shall consider, 
we need to know if ${\cal{O}}_{1,2} .  T_{r,1} . \int T_{r',1}$ is 
non--vanishing. Just by adding the Liouville momenta, and by minimal model 
fusion rules, we see that (schematically)
\be {\cal{O}}_{1,2} .  T_{r,1} . \int T_{r',1} \sim T_{r+r' -1,2}, \ee
for $2 \leq r +r' \leq q$. Explicitly, we get that
\bea \label{eqndet}
&& _{(m,n)}[ {\cal{O}}_{1,2}]_{(m,n \pm 1)} 
[ T_{r,1}]_{(m',n \pm 1)} \int [ T_{r',1}]_{(m'',n \pm 1)}  
\non \\
&&= F_{(m,n \pm 1)(r,2)} 
\left[ \begin{array}{cc}
(m,n) & (m' ,n \pm 1) \\
(1,2) & (r,1) 
\end{array}\right]
F_{(m',n \pm 1)(r+r' -1,2)} 
\left[ \begin{array}{cc}
(m,n) & (m'' ,n \pm 1) \\
(r,2) & (r',1) 
\end{array}\right] \non \\ && \times
_{(m,n)}[ T_{r+r' -1,2} ]_{(m'',n \pm 1)}.\eea  

The coefficient in \C{eqndet} is a product of fusion matrices because it is a 
four-point boundary correlator which equals the sum over intermediate 
channels of products of boundary three-point functions consistent 
with the fusion rules (in this case there is only one channel 
$(1,2) \times (r,1) \rightarrow (r,2)$).  
 
Similarly, because of the ordered nature of the integrated vertex 
operator, we have another relation (though we shall not need it, 
we list it for the sake of completeness)
\bea \label{eqndet2}
&& \int {_{(m,n)}[} T_{r,1}]_{(m',n)} [ {\cal{O}}_{1,2}]_{(m',n \pm 1)}
[ T_{r',1}]_{(m'',n \pm 1)}  
\non \\
&&= F_{(m',n)(r,2)} 
\left[ \begin{array}{cc}
(m,n) & (m' ,n \pm 1) \\
(r,1) & (1,2) 
\end{array}\right]
F_{(m',n \pm 1)(r+r' -1,2)} 
\left[ \begin{array}{cc}
(m,n) & (m'' ,n \pm 1) \\
(r,2) & (r',1) 
\end{array}\right] \non \\ && \times
_{(m,n)}[ T_{r+r' -1,2} ]_{(m'',n \pm 1)}.\eea  
Once again, the coefficient is given by a product of fusion 
matrices.\footnote{For $c_M =1$, the corresponding coefficients 
involve trigonometric functions and 
were derived in~\cite{Bershadsky:1992ub} for open string theory.
For closed string theory, the coefficient is given 
by the Virasoro-Shapiro four-point function and was derived
in~\cite{Klebanov:1991vp,Kachru:1992hw}.} It should be noted that
the coefficients in \C{eqndet} and \C{eqndet2} involve squares
of fusion matrices only when the matter part of the integrated 
vertex operator has a non-trivial OPE with the other two vertex 
operators. For example, consider the case when the integrated 
vertex operator is the boundary cosmological constant, which contains 
the identity in the matter part. Then in the matter sector 
there are no singularites that arise due to contact terms of the other 
two operators with the identity, and then the coefficients are 
different. In fact they are then given by matter three-point functions, 
and so they are linear (and not quadratic) in the fusion matrices. 

Now consider the case when $\mu_B$ is turned on, i.e. in
the presence of FZZT branes. We shall consider the equations involving
${\cal{O}}_{1,2}$, the results for the equations involving 
${\cal{O}}_{2,1}$ follow by replacing $b \rightarrow 1/b$, $\mu_B 
\rightarrow {\widetilde{\mu}}_B$, and suitably interchanging the Kac 
table indices. We want to see how \C{tachmod1} and \C{eqndet} are
modified in the presence of FZZT branes. The Liouville part of the
answer can be worked out along the lines of the previous section
as it involves inserting one boundary cosmological constant, and the minimal 
model part gives the fusion matrix. The first equation in \C{tachmod1} gets
modified to
\bea \label{needcal1}
&& _{(m,n)}^\s[ {\cal{O}}_{1,2}]^{\s \pm ib}_{(m,n \pm 1)} 
[ T_{r,1}]^{\s'}_{(m',n \pm 1)} = -{\sqrt \mu} 
\frac{r -b^2}{b^2{\rm sin} 2\pi b^2}  B(1+b^2,b^2 -r) \non \\
&&\times
\Big({\rm sin} 2\pi b^2 {\rm cosh} \pi b\s - {\rm sin} \pi b^2 \{
{\rm cosh} \pi b(\s \pm ib) -(-1)^r {\rm cosh} \pi b\s'\}\Big)
\non \\
&&\times
F_{(m,n \pm 1)(r,2)} 
\left[ \begin{array}{cc}
(m,n) & (m' ,n \pm 1) \\
(1,2) & (r,1) 
\end{array}\right] 
{_{(m,n)}^\s[} T_{r,2}]^{\s'}_{(m',n \pm 1)},
\eea
while the second equation in \C{tachmod1} gets modified to 
\bea \label{needcal2}
&& _{(m,n)}^\s[ {\cal{O}}_{1,2}]^{\s \pm ib}_{(m,n \pm 1)} [ T_{q-1,s}
]^{\s'}_{(m,n')} \non \\
&&= \frac{1-q +sb^2}{b^2} F_{(m,n \pm 1)(q-1,s-1)} 
\left[ \begin{array}{cc}
(m,n) & (m,n') \\
(1,2) & (q-1,s) 
\end{array}\right]
{_{(m,n)}^\s[} T_{q-1,s-1} ]^{\s'}_{(m,n')} \non \\
&& +{\sqrt \mu} \frac{1-q +sb^2}{b^2{\rm sin} (1+s) \pi b^2}
B(1+b^2,1-q +sb^2) \non \\
&& \times 
\Big({\rm sin} (1+s)\pi b^2 {\rm cosh} \pi b\s - {\rm sin} \pi s b^2  
{\rm cosh} \pi b(\s \pm ib) -(-1)^q {\rm sin} \pi b^2 {\rm cosh} 
\pi b\s'\Big) \non \\
&& \times F_{(m,n \pm 1)(q-1,s+1)} 
\left[ \begin{array}{cc}
(m,n) & (m,n') \\
(1,2) & (q-1,s) 
\end{array}\right] 
{_{(m,n)}^\s[} T_{q-1,s+1} ]^{\s'}_{(m,n')}. \eea

We have included the numerical factors that arise from the ghost,
$L_{-1}^M$ and $L_{-1}^L$ prefactors in \C{ringbasic}. This also leads
to a multiplicative factor of 
\be \frac{(r+r')(1+ b^2) -rr' -b^2 Q_L^2}{b^4}\ee
in \C{eqndet}.

Finally, we argue that \C{eqndet} does not receive any corrections. 
At $O(1)$, $T_{r+r' -1,2}$ emerges, as we saw before. At $O(\mu_B)$, 
by comparing Liouville momentum, $T_{r+r' -1,4}$ should emerge. Similarly,
at $O(\mu_B^2)$, $T_{r+r' -1,6}$ 
should emerge, and so on. However, from the left-hand side of \C{eqndet},
using the minimal model fusion rules, we see that only tachyons of the form
$T_{k,2}$ can emerge. Hence, all terms except the $O(1)$ term vanish. It
is also easy to see the vanishing in terms of the fusion matrix. For 
example, at $O(\mu_B)$, this is due to the factor of
\be  F_{(m',n \pm 1)(r+r' -1,4)} 
\left[ \begin{array}{cc}
(m,n) & (m'',n +1) \\
(r,2) & (r',1) 
\end{array}\right] =0,\ee
which enters the matter four-point function. 

We now deduce a
difference equation satisfied by boundary correlators. Generic correlators 
of the form 
\be \langle ^{\s_1}_{(m_1,n_1)}[ T_{r_1,s_1}]  ^{\s_2}_{(m_2,n_2)}[ 
T_{r_2,s_2}]^{\s_3}_{(m_3,n_3)}  \ldots 
{^{\s_k}_{(m_k,n_k)}[} T_{r_k,s_k}]^{\s_1}_{(m_1,n_1)}  \rangle \ee
involving only boundary tachyons are cumbersome to manipulate. So we 
consider a subclass of the most general correlators which are 
tractable. The arguments we present go through for the general case as 
well. We consider the correlator
\be \label{calcorr}
\langle ^{\s}_{(m,n)}[ T_{q-1,s}]  ^{\s'}_{(m,n')}[ 
T_{r,1}] ^{\s''}_{(m',n')}[ T_{r',1}]^{\s'''}_{(m'',n')}  \ldots 
\rangle,\ee
where we have not distinguished between integrated 
and unintegrated vertex operators
for brevity -- it should be kept in mind 
that three of the vertex operators are 
not integrated while the rest are integrated over in a specific manner 
as we now discuss. The first three tachyons have been chosen to be of 
this form so that we can use the ring relations acting on the tachyon 
module given by \C{eqndet}, \C{needcal1} and \C{needcal2}, while the
remaining tachyons are arbitrary. The tachyons $T_{q-1,s}$ and $T_{r,1}$ are
at 0 and 1 respectively, while the last tachyon in \C{calcorr} is
at $\infty$. Also the integrated tachyons in \C{calcorr} are 
ordered along the boundary and not integrated over the whole 
boundary, in particular, $T_{r',1}$ is an ordered integrated vertex operator. 
Thus this correlator gives a particular channel of the scattering 
amplitude.       
 
Consider the correlator 
\be \langle ^{\s}_{(m,n)}[ T_{q-1,s}]  ^{\s'}_{(m,n')}[ 
{\cal{O}}_{1,2}]^{\s' \pm ib}_{(m,n' \pm 1)}
[ T_{r,1}] ^{\s''}_{(m',n' \pm 1)}[ T_{r',1}]^{\s'''}_{(m'',n' \pm 1)}
\ldots \rangle,\ee
where the various tachyons are inserted on the worldsheet as mentioned 
above. For this particular choice of insertions, this correlator is
independent of the insertion point of 
${\cal{O}}_{1,2}$~\cite{Bershadsky:1992ub}. We evaluate this correlator
in two ways: by allowing $T_{q-1,s} \leftarrow {\cal{O}}_{1,2}$, and 
${\cal{O}}_{1,2} \rightarrow T_{r,1} \int T_{r',1}$, and equate the answers,
on using \C{eqndet}, \C{needcal1} and \C{needcal2}. This gives us
the following relation among correlators
\bea \label{eqnamongcor}
&& \alpha {\widetilde{M}}_1
\langle 
^\s_{(m,n)}[ T_{q-1,s} ] ^{\s'}_{(m,n')}[ T_{r+r' -1,2} 
]^{\s'''}_{(m'',n' \pm 1)}
\ldots \rangle \non \\
&& -{\sqrt \mu} 
\beta f(\s,\s') {\widetilde{M}}_2
\langle ^\s_{(m,n)}[ T_{q-1,s} ]
{_{(m,n')}^{\s'}[} T_{r,2}]^{\s''}_{(m',n' \pm 1)} 
[ T_{r',1}]^{\s'''}_{(m'',n' \pm 1)} \ldots \rangle \non \\
&& =\gamma {\widetilde{M}}_3
\langle {_{(m,n)}^\s[} T_{q-1,s-1} ]^{\s' \pm ib}_{(m,n' \pm 1)} 
[ T_{r,1}]^{\s''}_{(m',n' \pm 1)}
[ T_{r',1}]^{\s'''}_{(m'',n' \pm 1)} \ldots \rangle \non \\
&& -{\sqrt \mu} \tau g (\s,\s') {\widetilde{M}}_4
\langle {_{(m,n)}^\s[} T_{q-1,s+1} ]^{\s' \pm ib}_{(m,n' \pm 1)}
[ T_{r,1} ]^{\s''}_{(m',n' \pm 1)} [ T_{r',1} ]^{\s'''}_{(m'',n' \pm 1)} 
\ldots \rangle, \eea
where
\bea \label{longrel}
&& \alpha = \frac{(r+r')(1+ b^2) -rr' -b^2 Q_L^2}{b^4}, \quad \beta 
=\frac{r -b^2}{b^2{\rm sin} 2\pi b^2}  B(1+b^2,b^2 -r), \non \\
&& \gamma =\frac{q-1 -sb^2}{b^2} ,\qquad
\tau =-\frac{\gamma}{{\rm sin} (1+s) \pi b^2}
B(1+b^2,1-q +sb^2), \non \\
&& f (\s,\s') = {\rm sin} 2\pi b^2 {\rm cosh} \pi b\s' - {\rm sin} \pi b^2 \{
{\rm cosh} \pi b(\s' \pm ib) -(-1)^r {\rm cosh} \pi b\s''\}, \non \\
&& g (\s,\s') ={\rm sin} (1+s)\pi b^2 {\rm cosh} \pi b(\s' \pm ib) - 
{\rm sin} \pi s b^2 {\rm cosh} \pi b\s'  
-(-1)^q {\rm sin} \pi b^2 {\rm cosh} 
\pi b\s, \non \\
&& {\widetilde{M}}_1 =F_{(m,n' \pm 1)(r,2)} 
\left[ \begin{array}{cc}
(m,n') & (m' ,n' \pm 1) \\
(1,2) & (r,1) 
\end{array}\right] 
F_{(m',n' \pm 1)(r+r' -1,2)} 
\left[ \begin{array}{cc}
(m,n') & (m'' ,n' \pm 1) \\
(r,2) & (r',1) 
\end{array}\right] , \non \\
&& {\widetilde{M}}_2 =F_{(m,n' \pm 1)(r,2)} 
\left[ \begin{array}{cc}
(m,n') & (m' ,n' \pm 1) \\
(1,2) & (r,1) 
\end{array}\right] , \non \\
&& {\widetilde{M}}_3 =F_{(m,n')(q-1,s-1)} 
\left[ \begin{array}{cc}
(m,n) & (m,n' \pm 1) \\
(q-1,s) & (1,2) 
\end{array}\right] ,\non \\
&& {\widetilde{M}}_4=
F_{(m,n')(q-1,s+1)} 
\left[ \begin{array}{cc}
(m,n) & (m,n' \pm 1) \\
(q-1,s) & (1,2) 
\end{array}\right]. \eea

We want to write down an equation satisfied by \C{calcorr} starting 
from \C{eqnamongcor}. In \C{eqnamongcor} the first term 
is a correlator with one less tachyon than the others. Also from 
\C{longrel} we see that all the other terms depend on both  
$\s' \pm ib$, while the first term in \C{eqnamongcor} does not. So we
take the difference between the two equations \C{eqnamongcor}
involving $\s' + ib$ and $\s' -ib$ and thus the first terms drops out. 
As we now demonstrate, this gives a formal equation satisfied by the 
correlator \C{calcorr}. 

In order to write down the equation, we use the identity 
in the Liouville sector~\cite{Kostov:2003cy} 
\be 
\langle ^\s[T] ^{\s' \pm ib}[T] ^{\s''}[T] \ldots \rangle
=e^{\pm b \p_{\s'}} \langle ^\s[T] ^{\s'}[T] ^{\s''}[T] 
\ldots \rangle. \ee
We need an analogous relation in the matter sector, were the variables are
discrete as they lie in the Kac table, and so we define a formal 
differentiation to shift the indices. As we shall see, this definition
is natural given the fusion rule constraint. For the cases we need, 
we define
\bea && \langle _{(m,n)}[T_{q-1,s}] _{(m,n')}[T_{r,2}] 
_{(m',n' \pm 1)}[T_{r',1}]_{(m'',n' \pm 1)} \ldots \rangle \non \\
&& =e^{-i\frac{b}{2} \p_{r,1}} \langle _{(m,n)}[T_{q-1,s}] _{(m,n')}[T_{r,1}]
_{(m',n')}[T_{r',1}]_{(m'',n')} \ldots \rangle. \eea

The exponential factor adds momentum $b/2$ to $T_{r,1}$, and 
since $\alpha_{r,1} +b/2 =\alpha_{r,2}$, we get $T_{r,2}$. Also the 
fusion rules between $(m,n')$ and $(r,2)$ automatically change the 
remaining index at that insertion point from $(m',n')$ to $(m',n'
\pm 1)$, which changes the index from $(m'',n')$ to $(m'',n' \pm 1)$ 
across $T_{r',1}$, and so on. Similarly it follows that
\bea && \langle _{(m,n)}[T_{q-1,s + 1}] _{(m,n' \pm 1)}[T_{r,1}] 
_{(m',n' \pm 1)}[T_{r',1}]_{(m'',n' \pm 1)} \ldots \rangle \non \\
&& =e^{-i\frac{b}{2} \p_{q-1,s}} \langle _{(m,n)}[T_{q-1,s}] 
_{(m,n')}[T_{r,1}] _{(m',n')}[T_{r',1}]_{(m'',n')} \ldots \rangle,\eea
and
\bea && \langle _{(m,n)}[T_{q-1,s -1}] _{(m,n' \pm 1)}[T_{r,1}] 
_{(m',n' \pm 1)}[T_{r',1}]_{(m'',n' \pm 1)} \ldots \rangle \non \\
&& =e^{i\frac{b}{2} \p_{q-1,s}} \langle _{(m,n)}[T_{q-1,s}] 
_{(m,n')}[T_{r,1}] _{(m',n')}[T_{r',1}]_{(m'',n')} \ldots \rangle.\eea

Using these definitions, we get the difference equation satisfied by 
\C{calcorr}
\bea && \Big( \gamma {\widetilde{M}}_3 {\rm sinh} (b \p_{\s'}) 
e^{i\frac{b}{2} \p_{q-1,s}} -i \sqrt{\mu} \beta {\rm sin}^2 (\pi b^2)
{\rm sinh} (\pi b\s') {\widetilde{M}}_2 e^{-i\frac{b}{2} \p_{r,1}} 
\non \\
&& -\sqrt{\mu} \tau {\rm sin} (\pi b^2) {\widetilde{M}}_4 \{ h (\s,\s')
{\rm sinh} (b \p_{\s'}) +i u (\s,\s') {\rm cosh} (b \p_{\s'}) \}
e^{-i\frac{b}{2} \p_{q-1,s}} \Big) \non \\
&& \times \langle ^{\s}_{(m,n)}[ T_{q-1,s}]  ^{\s'}_{(m,n')}[ 
T_{r,1}] ^{\s''}_{(m',n')}[ T_{r',1}]^{\s'''}_{(m'',n')}  \ldots 
\rangle =0,\eea
where
\bea 
&& h (\s,\s') = {\rm cosh} \pi b\s' {\rm cos} \pi b^2 (1+s)
-(-1)^q {\rm cosh} \pi b\s, \non \\
&& u (\s,\s') = {\rm sinh} \pi b\s' {\rm sin} \pi b^2 (1+s).
\eea

\section{Discussion}

In this work we have focussed exclusively on worldsheet methods to obtain 
the various results. Naturally one would like to derive them using the 
dual matrix models -- in particular, in order to go beyond the 
perturbative definition of minimal open string theory. The matrix models
in question here are the hermitian two-matrix models, with polynomial
potentials of degree $p$ and $q$ in the two matrices $X$ and $Y$
respectively~\cite{Daul:1993bg}.  
%In fact the exact answers can be 
%very different from the perturbative ones. 

A Riemann surface ${\cal{M}}_{p,q}$ arises in describing the 
classical target space of minimal string theory as the moduli 
space of FZZT branes~\cite{Seiberg:2003nm}.  This Riemann surface has a 
complicated sheeted structure determined by the values of $p$ and $q$. 
The coordinates $x$ and $y$ parametrizing the eigenvalue planes 
of $X$ and $Y$ can be identified with $\mu_B$ and its dual $\tilde\mu_B$
under $b\to 1/b$.
The space ${\cal{M}}_{p,q}$ is a $q$-sheeted cover of the $x$ plane
and a $p$-sheeted cover of the $y$ plane;
$\sigma$ is a uniformizing coordinate for the surface.
Asymptotically, for large $x$, $y$, the shifts
$\sigma\to\sigma+ib$ and $\sigma\to\sigma+i/b$ are deck
transformations taking one between the various sheets.
Thus the relation \C{lowmat} equating matter boundary
states at fixed $\sigma$ to a sum over states at different values
of $\sigma$ tells us that
information about the matter states is encoded nonlocally
on the surface ${\cal{M}}_{p,q}$ (although for example
$|\sigma;1,k\rangle$ lives on the points over a single value of $y$,
and $|\sigma;j,1\rangle$ sits over a single $x$).
Microscopically, minimal models are described as 
lattice statistical models with degrees of freedom
taking values in a Dynkin diagram \cite{DiFrancesco:1997nk}.
The matter boundary conditions $(1,k)$ describe 
a Dirichlet boundary condition at a fixed location
on the Dynkin diagram \cite{Saleur:1988zx},
which thus appears to be encoded in the sheets of ${\cal{M}}_{p,q}$.
Interestingly, this structure seems to be deformed,
and disappear altogether, as one passes to the strong coupling
region at small $x$, $y$.
Moreover, as argued in~\cite{Maldacena:2004sn},
the quantum target space turns out to be very 
different since the FZZT brane correlators 
are entire functions of $\mu_B$.  The quantum target space 
reduces to just the complex plane, and the structure of
the Dynkin diagram is less apparent. 

It would be helpful to deduce the various results above using matrix models:
for example, the relation among boundary states \C{eqstate}. An 
understanding of the boundary ground ring in matrix models is definitely
going to be useful in elucidating the relation between
the worldsheet formulation and the nonperturbative
formulation given by the matrix model. Also in evaluating
correlators in matrix models, one has to know how to encode the matter 
boundary conditions.  The boundary states of the worldsheet
theory are labelled by both $\sigma$, which parametrizes 
the eigenvalue plane of the matrices; and by the matter
boundary conditions $(m,n)$.  So far, there is no direct
understanding of these labels in the matrix model.

It is conceivable that the matter boundary conditions might be related to 
the choice of the contours of integration on the eigenvalue planes. One can 
deform the contours to start and end at infinity in such a way that the 
integral converges. There exists such a choice of 
contours~\cite{Bertola:2001hq,Bertola:2001hq2} which splits the 
eigenvalue planes into angular sectors. In fact there are 
$p-1$ and $q-1$ independent such contours in the $x$ and $y$ 
planes respectively: the same number as the choices of 
$(1,s)$ and $(r,1)$ matter boundary conditions. Considering the 
Hilbert-Laplace transform of operators along these 
contours~\cite{Bertola:2001hq2} might be related to the choice of matter 
boundary conditions. A change of basis from these contours to 
the steepest descent contours as mentioned in~\cite{Bertola:2001hq2} may
be a natural choice for specifying the matter boundary conditions. 
It would be interesting to understand concretely how the matter 
boundary conditions arise from the two-matrix model. 

Another issue which we have not addressed concerns the structure
of the boundary ground ring itself.  The bulk ground ring is
generated by the bulk $\Co_{1,2}$ and $\Co_{2,1}$ operators,
subject to the relations
\be
U_{q-1}({\cal{O}}_{1,2})=U_{p-1}({\cal{O}}_{2,1})=0\ .
%U_{q-2}(\Co_{1,2})-U_{p-2}(\Co_{2,1})=0
\ee
The boundary ground ring is similarly generated by
the boundary $\Co_{1,2}$ and $\Co_{2,1}$ operators,
since by fusion these operators generate any $\Co_{r,s}$
in their products; however, the ring algebra 
and relations have not been worked out in this case.
The multiplication table is not as simple as in the bulk
case; for instance, the product of $\Co_{1,2}$ with itself
to make the identity depends only on $\mu_B$,
while the product of $\Co_{1,3}$ with itself to
make the identity depends on both $\mu$ and $\mu_B$.
The fact that the multiplication table depends
on $\mu_B$ means that the whole structure varies
as one moves around on the Riemann surface ${\cal{M}}_{p,q}$.
Furthermore, the boundary ground ring operators are 
endomorphisms in the space of open string boundary conditions.  
Thus there is room for non-commutative structure,
and indeed the product $\Co_{1,2}\Co_{2,1}$ differs
from the product $\Co_{2,1}\Co_{1,2}$ since they are
determined by distinct fusion coefficients.
We leave the characterization of the boundary ground ring
in terms of generators, relations, {\it etc.},  to future work.
   
\section*{Acknowledgements}

The work of A.~B. is supported in part by  NSF Grant No.
PHY-0204608, and the work of E.~M. is supported in part by
DOE grant DE-FG02-90ER40560. 

%\newpage
%\bibliographystyle{amsunsrt-es}
%\bibliography{myrefs}
%\bibliographystyle{utphys}
%\bibliography{myrefs}

%\end{thebibliography}
%\end{document}

%\begin{thebibliography}{10}

%\ifx\undefined\bysame
%\newcommand{\bysame}{\leavevmode\hbox to3em{\hrulefill}\,}
%\fi

\providecommand{\href}[2]{#2}\begingroup\raggedright\endgroup

%\end{thebibliography} 
\end{document}